\documentclass[10pt,preprint2]{aastex}

\def\gtsim{\raise 2pt \hbox {$>$} \kern-1.1em \lower 4pt \hbox {$\sim$}}
\def\ltsim{\raise 2pt \hbox {$<$} \kern-1.1em \lower 4pt \hbox {$\sim$}} 

\shorttitle{Parsec Scale Properties of Markarian 501}
\shortauthors{Giroletti et al.}

\begin{document}

\title{Parsec Scale Properties of Markarian 501} 

\author{M. Giroletti\altaffilmark{1,2},
        G. Giovannini\altaffilmark{1,2}, 
        L. Feretti\altaffilmark{1},
        W.D. Cotton\altaffilmark{3}, 
        P.G. Edwards\altaffilmark{4},
        L. Lara\altaffilmark{5,6}, 
        A.P. Marscher\altaffilmark{7},
        J.R. Mattox\altaffilmark{8}, 
        B.G. Piner\altaffilmark{9},
        T. Venturi\altaffilmark{1}      }


\altaffiltext{1}{Istituto di Radioastronomia del CNR, via Gobetti 101,
40129, Bologna, Italy}

\altaffiltext{2}{Dipartimento di Astronomia, Universit\`a di Bologna,
via Ranzani 1, 40127 Bologna, Italy}

\altaffiltext{3}{National Radio Astronomy Observatory, 520 Edgemont
Road, Charlottesville, VA 22903-2475, USA}

\altaffiltext{4}{Institute of Space and Astronautical Science, 3-1-1
Yoshinodai, Sagamihara, Kanagawa 229-8510, Japan}

\altaffiltext{5}{Dpto. Fisica Teorica y del Cosmos, Universidad de
Granada, 18071 Granada, Spain}

\altaffiltext{6}{Instituto de Astrofisica de Andalucia (CSIC),
Apdo. 3004, 18080 Granada, Spain}

\altaffiltext{7}{Institute for Astrophysical Research, Boston
University, 725 Commonwealth Ave., Boston, MA 02215, USA}

\altaffiltext{8}{Department of Physics and Astronomy, Francis Marion
University, Florence, SC 29501-0547, USA}

\altaffiltext{9}{Department of Physics and Astronomy, Whittier
College, 13406 East Philadelphia Street, Whittier, CA 90608, USA}


\begin{abstract}

We present the results of a high angular resolution study of the BL
Lac object Markarian 501 in the radio band.  We consider data taken at
14 different epochs, ranging between 1.6\,GHz and 22\,GHz in frequency,
and including new Space VLBI observations obtained on 2001 March 5 and
6 at 1.6 and 5\,GHz. We study the kinematics of the parsec-scale jet
and estimate its bulk velocity and orientation with respect to the
line of sight.  Limb brightened structure in the jet is clearly
visible in our data and we discuss its possible origin in terms of
velocity gradients in the jet.  Quasi-simultaneous multi-wavelength
observations allow us to map the spectral index distribution and to
compare it to the jet morphology.  Finally, we estimate the 
physical parameters of the parsec-scale jet.

\end{abstract}

\section{INTRODUCTION}

BL Lac objects are one of the several flavors of radio loud active
galactic nuclei. Along with peculiar properties at other wavelengths
(lack of strong emission lines, high levels of variability, optical
polarization up to 3\%, detection at X-ray and $\gamma$-ray energies),
they also show quite extreme behavior in the radio: the parsec-scale
structure is dominated by a compact, flat-spectrum core, from which a
one-sided jet emerges, showing knots of enhanced brightness. When
monitored over long intervals, these components are often found to be
in motion, sometimes with an apparently superluminal velocity
\citep[see, e.g.,][]{hom01}.

The currently accepted explanation for these properties is found in
the frame of unified models \citep[see e.g.][]{urr95}. These models
are based on a dust enshrouded super-massive black hole accreting
matter in a disk, from which two symmetric collimated jets of
relativistic plasma are ejected in opposite directions. When the angle
between the jet axis and the line of sight is small, the resulting
Doppler boosting accounts for many of the above-mentioned properties.

A source such as Markarian 501 (B1652+39) is an ideal target to test
the assumptions involved in unified models. At its low redshift
($z=0.034$)\footnote{We assume H$_0=65$\,km\,s$^{-1}$\,Mpc$^{-1}$},
1\,mas~=~0.72\,pc; therefore, the milliarcsecond resolution achievable
with VLBI techniques makes it possible to investigate in great detail
the regions near the central core. Furthermore, Mkn~501 is one of the
few sources with a clear limb-brightened structure
\citep{gio99,aar99}.  Finally, this source is very well studied at
other frequencies and it is known for its X$-$ and $\gamma-$ray
\citep{qui96,bra97} activity, which also sets constraints on the
parameters describing the physics of the inner jet \citep[see][and
references therein]{tav01,kat01}.

In this paper we present a large multi-epoch, multi-frequency data-set
of VLBI images; new observations as well as previously published data
are used to study the physical properties of this source. In
particular, we will discuss the pc-scale jet morphology, velocity and
orientation.  Additionally, our data-set allows us to compare images
obtained at different frequencies with similar resolutions and thus to
perform spectral index\footnote{We will define the spectral index
$\alpha$ such that $S(\nu) \propto \nu^{- \alpha}$} studies for
comparison with the total intensity images.

In \S\,\ref{sec:observations} we describe the previous and new
observations used in this paper, along with the data reduction methods
and procedures for component fitting and spectral index
mapping. Results are presented in \S\,\ref{sec:results} and discussed
in \S\,\ref{sec:discussion}. We present our conclusions in
\S\,\ref{sec:conclusions}.

\section{OBSERVATIONS}
\label{sec:observations}

Table 1 summarizes all observations considered in this paper, listing
epoch, frequency, total observing time and array used. The notes
provide references for details.  The present data-set is very similar
to the one used by Edwards \& Piner (2002, hereafter
\citeauthor{edw02}) to discuss the proper motion in Mkn 501: we did
not use the 1995 Oct 17 epoch (due to its low sensitivity), or the
1996 Jul 10 and 1998 Oct 30 epochs (as better data from other nearby
epochs was available).  On the other hand, we add 3 more epochs (see
note 4 to Table 1) at 15 and 22\,GHz and we consider two new Space
VLBI (VSOP) observations obtained on 2001 March 5 and 6 (see Table~2).

\subsection{Data Reduction}

\subsubsection{Ground Observations}

Images from some of the previous observations have been published or
made available to the scientific community on open web sites. However,
for the sake of completeness and homogeneity, we requested and
obtained the calibrated ($u, v$)-data for all observations presented
here. After the initial reduction (see notes to Table~1 for
references), we imported the calibrated data into the NRAO
Astronomical Image Processing System (AIPS) and inspected them
carefully. Then, we produced new images; one or more self-calibration
cycles proved to be useful in some cases, providing a better signal to
noise ratio. We adopted the appropriate ($u, v$)-range, weights,
cell-size and angular resolution to map the spectral index and to look
for proper motion of the jet substructures; we present parameters of
final images in Table~3.

\subsubsection{Space VLBI Observations}

We obtained Space VLBI images at three different epochs: 1997.8,
1998.4, and 2001.3.  VSOP (VLBI Space Observatory Programme)
observations combine data from ground arrays, such as the VLBA, and
the 8\,m orbiting antenna on board the satellite HALCA. For the latter
two epochs observations at 1.6 and 4.8\,GHz were performed on
successive days to minimize flux density variability problems in
spectral index mapping.  The first Space VLBI observation was
conducted at 1.6\,GHz only.  Further details on these observations are
given in Table~2.

The standard VSOP observing mode has one polarization (LCP) and two
IFs of 16 MHz each with two-bit sampled data.  The lowest IF failed in
the 1.6\,GHz observation in 1997.  In 2001, the ``a priori''
calibration was affected by lack of system temperature measurement for
Robledo and Goldstone. A mini-cal procedure was done before starting
the experiment in Robledo, providing a value of 20\,K that was assumed
as constant. For HALCA, we used the nominal system temperatures of
75\,K and 77\,K for the two IFs at 1.6\,GHz, and 88\,K and 92\,K for
those at 4.8\,GHz \citep{hir98,hir00}.
 
All the data were correlated in Socorro and then imported and reduced
in AIPS. Global fringe-fitting returned good solutions for all the
observations in which large ground telescopes were used, although no
solutions were found for the last observation (2001 Mar 6), unless a
value of SNR as low as 4.5 was set in the fit. In spite of such a low
signal to noise ratio, we found these solutions reasonable: delays and
rates found in this way do not present any discontinuity or suspicious
behavior. Thus, we applied these solutions and proceeded with the
whole array.

\begin{figure}
\figcaption{($u, v$)-coverage of the Space VLBI observation at 1.6\,GHz
($\lambda=18$~cm) on 1998 April 8. \label{uv1} {\it Available
    separately as .jpg file 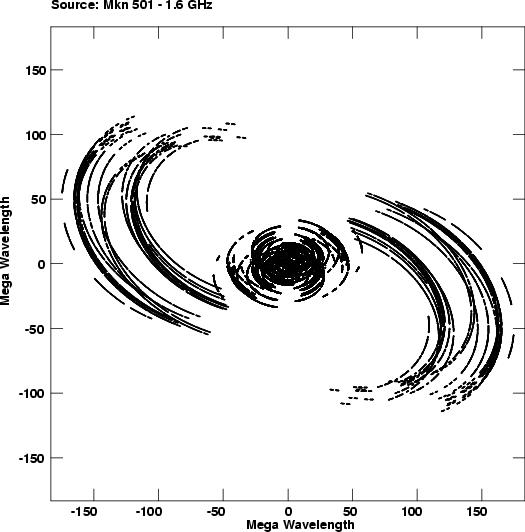} }
\end{figure}

\begin{figure}
\figcaption{($u, v$)-coverage of the Space VLBI observation at 4.8\,GHz
($\lambda=6$~cm) on 1998 April 7. \label{uv2} {\it Available
    separately as .jpg file 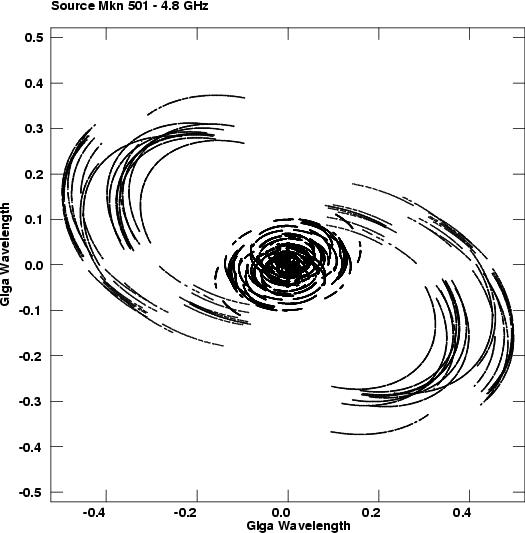}}
\end{figure}

We then performed self calibration using clean-component models, first
in phase and, afterward, in both phase and amplitude. In the latter,
we repeated the task several times, increasing the baseline
range. This revealed to be of great importance for those observations
where gain information for some antennas was missing. For example, we
could self-calibrate the shortest space-ground baselines using
`uniform weight' ground-only images; in fact, we have similar
baselines both in the ground array and to HALCA, as provided by its
highly elliptical orbit (see Fig.~\ref{uv1}, \ref{uv2}). Figures
\ref{fig3} -- \ref{fig9} present all the images, in order of
decreasing angular resolution.

\subsection{Component Fitting}

Identification of components is usually the most difficult problem
affecting proper motion studies. Long time intervals between epochs
can lead to misidentification of components; moreover, the use of
different resolutions can cause confusion, for example by blending
components with different properties and velocities; finally,
registering images at different observing frequencies is also
necessary: the apparent core shift due to the different spectral index
of the core and jet can be significant.

In the present work, the time coverage is very dense (14 epochs over 4
years, plus the final VSOP observation) and there are no large time
gaps. To prevent angular resolution problems, we produced maps at the
same resolution at all epochs. Even if some observations allowed a
better resolution, we convolved all the images obtained at a frequency
5\,GHz\,$\le \nu \le$\,15\,GHz with a circular restoring beam of
1.2\,mas FWHM. However, higher resolution maps have also been made, in
order to provide more information when needed. In particular we
obtained images at a resolution of 0.6 $\times$ 0.9 mas (RA $\times$
Dec) in all observations with good $(u,v)$-coverage on long baselines
(see Table~3).  A comparison of fit components in convolved and full
resolution images revealed good agreement.

The fit to the components of the jet was done within AIPS and the
Difmap package\footnote{Difmap was written by Martin Shepherd at
Caltech and is part of the VLBI Caltech Software package}.  In AIPS we
used the task \texttt{JMFIT}, which allows one to fit Gaussian
components to part of an image. Up to four components can be used at a
time and the task, starting from given initial conditions, solves in
an iterative way for position, flux density and extension of each
component.  Since more than four components are needed to properly fit
the images, we had to run the task several times on different
regions. The regions were made to overlap with each other in order to
obtain results compatible with previously published models.  We also
repeated the fit looking for ``stable'' solutions, i.e., changing our
initial guesses or even leaving them blank. Eventually, very good
models were obtained; the difference in flux density between our fits
and the images is always smaller than 5\%.

Using Difmap, we also carried out model-fitting to the visibilities
rather than the image. Similarly to the fits made in AIPS, four to six
jet components were required in addition to the core to provide a good
representation of the data.

\subsection{Spectral Index Mapping}

When producing spectral index images, there are three main effects
that have to be carefully taken into account. First, because of flux
density variability and possible proper motion of components, the
observations must be taken at very close epochs. Next, the angular
resolution must be the same at both frequencies and the
$(u,v)$-coverage as similar as possible; this guarantees the same
sensitivity to extended structures in the high and low frequency
images. Finally, as absolute position is not available, we need to
properly align the images before combining them.

To meet the first two requirements, we produced spectral index images
using data separated by less than 24\,hrs. We considered 15 and
22\,GHz data obtained on the same day (1997 Aug 15) and Space VLBI
observations at 1.6 and 5\,GHz made on successive days \citep[1998 Apr
7 and 8 --- see][for a preliminary analysis]{edw00}.  We note that
1998 Space VLBI observations have better $(u,v)$-coverage than those
in 2001 (see Fig.~\ref{uv1}, \ref{uv2}). To obtain matching
$(u,v)$-coverages, we cut the shortest baselines from the low
frequency data and the longest baselines in the high frequency
observations. Images were then obtained using uniform weights, and the
same cell-size and restoring beam.

The problem in registering the images is that the peak of the image,
coincident with the core, can correspond to different positions at
different frequencies if the core is self-absorbed.  So, we aligned
the images using the positions of components along the jet in a region
with steep spectral index and no self-absorption effects. We found
that at 15 and 22\,GHz no shift was necessary, while a shift of
$\sim$0.2\,mas was necessary in both RA and Dec between the 5 and
1.6\,GHz images. See Figures \ref{spix1} -- \ref{spix3} for the final
spectral index maps.

\section{RESULTS}
\label{sec:results}

\subsection{Source Morphology}
\label{sec:morphology}

\begin{figure}
\plotone{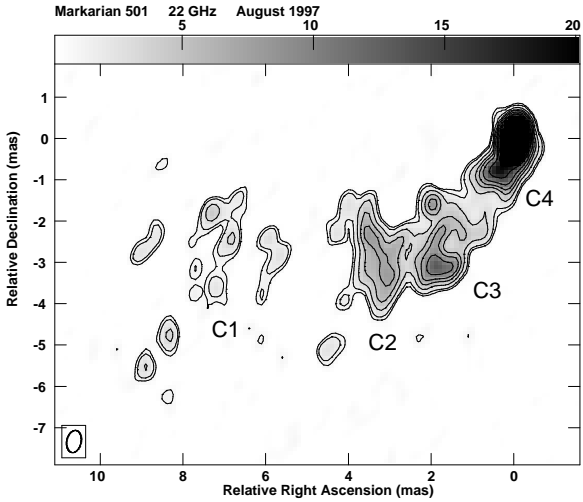}
\figcaption{Isocontour level of Mkn 501 at 22\,GHz from the August 1997 epoch. 
The HPBW is 0.54\,$\times$\,0.35\,mas in PA $-11^\circ$.
The noise level is 0.3\,mJy/beam. Contours are drawn at 1, 1.5 3, 5, 7, 10, 15,
20, 30, 50, 70, 100, 200, and 300 mJy/beam; the greyscale range is 
0.3--20\,mJy/beam. The limb brightened structure is visible even in the
innermost region of the jet. \label{fig3}}
\end{figure}

\begin{figure}
\plotone{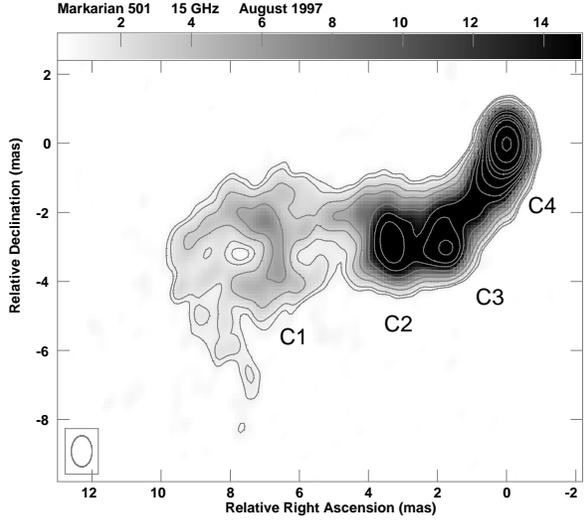}
\figcaption{Isocontour level of Mkn 501 at 15\,GHz from the August 1997 epoch.  
The HPBW is $0.6 \times 0.9$\,mas (RA\,$\times$\,Dec).  The noise level is 
0.2\,mJy/beam. Contours are drawn at 1, 1.5 3, 5, 10, 15, 20, 30, 50, 70, 100,
200, and 400\,mJy/beam; the greyscale range is 0.3--15\,mJy/beam. The limb
brightening is clearly visible in the (resolved) C1 region, but no longer in
the inner region because of the lower resolution. \label{fig4}}
\end{figure}

\begin{figure}
\plotone{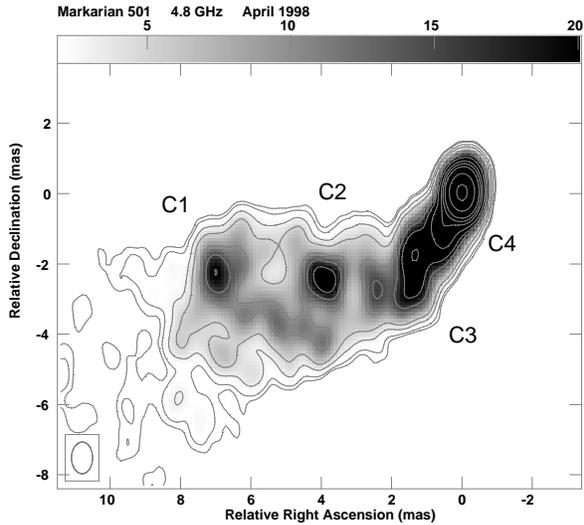}
\figcaption{Gray scale plus isocontour level of Mkn 501 from the 5\,GHz VSOP
observation in April 1998.  The HPBW is $0.6 \times 0.9$\,mas 
(RA\,$\times$\,Dec). 
Contours are drawn at 1, 1.5 3, 5, 10, 15, 20, 30, 50, 70, 100, 200,
and  400\,mJy/beam; the greyscale range is 2--20\,mJy/beam. \label{fig5}}
\end{figure}

\begin{figure}
\plotone{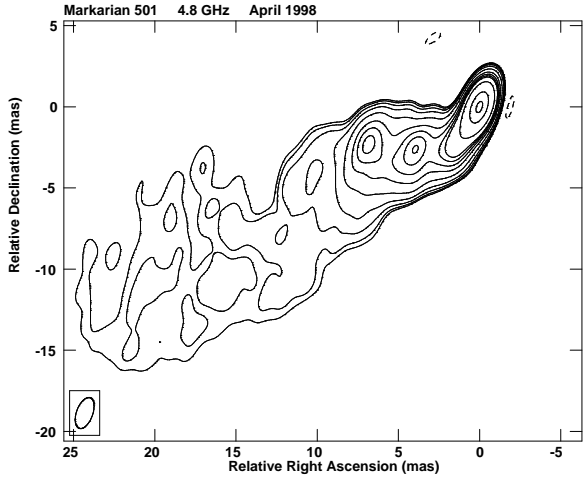}
\figcaption{Isocontour level of Mkn 501 from the 5\,GHz VSOP 
observation in April 1998. The
HPBW is 2\,$\times$\,1\,mas in PA $-20^\circ$. Contours are drawn at $-$3, 1.5,
2, 3, 5, 10, 20, 30, 40, 50, 100, 300, and 500\,mJy/beam. \label{fig6}}
\end{figure}

\begin{figure}
\plotone{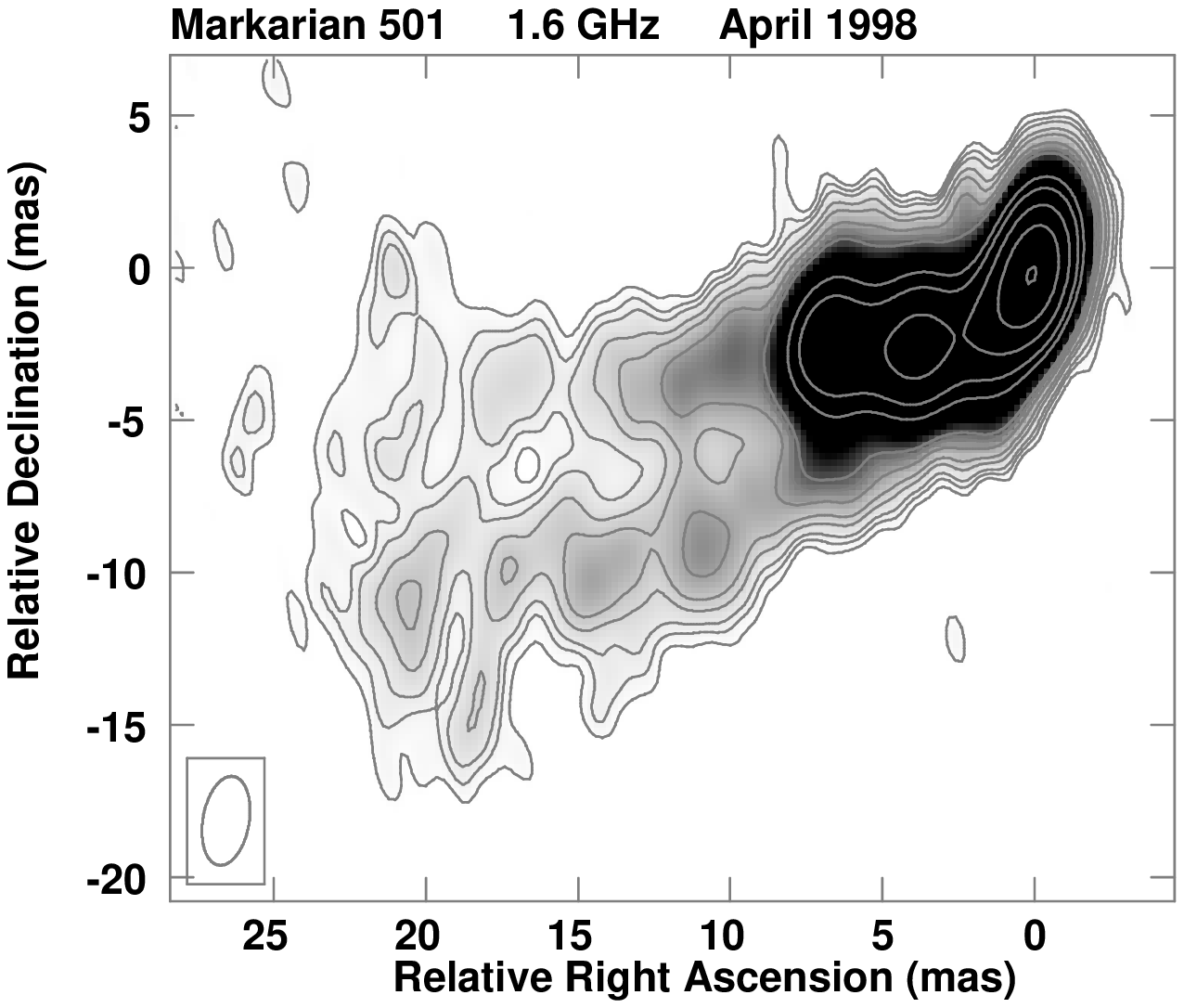}
\figcaption{Gray scale plus isocontour level of Mkn 501 from the 1.6\,GHz VSOP
observation in April 1998. The HPBW is 3\,$\times$\,1.5\,mas at 
PA $-10^\circ$. Contours are
drawn at 1, 1.5, 2, 3, 4, 6, 8, 10, 30, 50, 100, 200, and 400\,mJy/beam;
the greyscale range is 0.80--15\,mJy/beam. \label{fig7}}
\end{figure}

\begin{figure}
\plotone{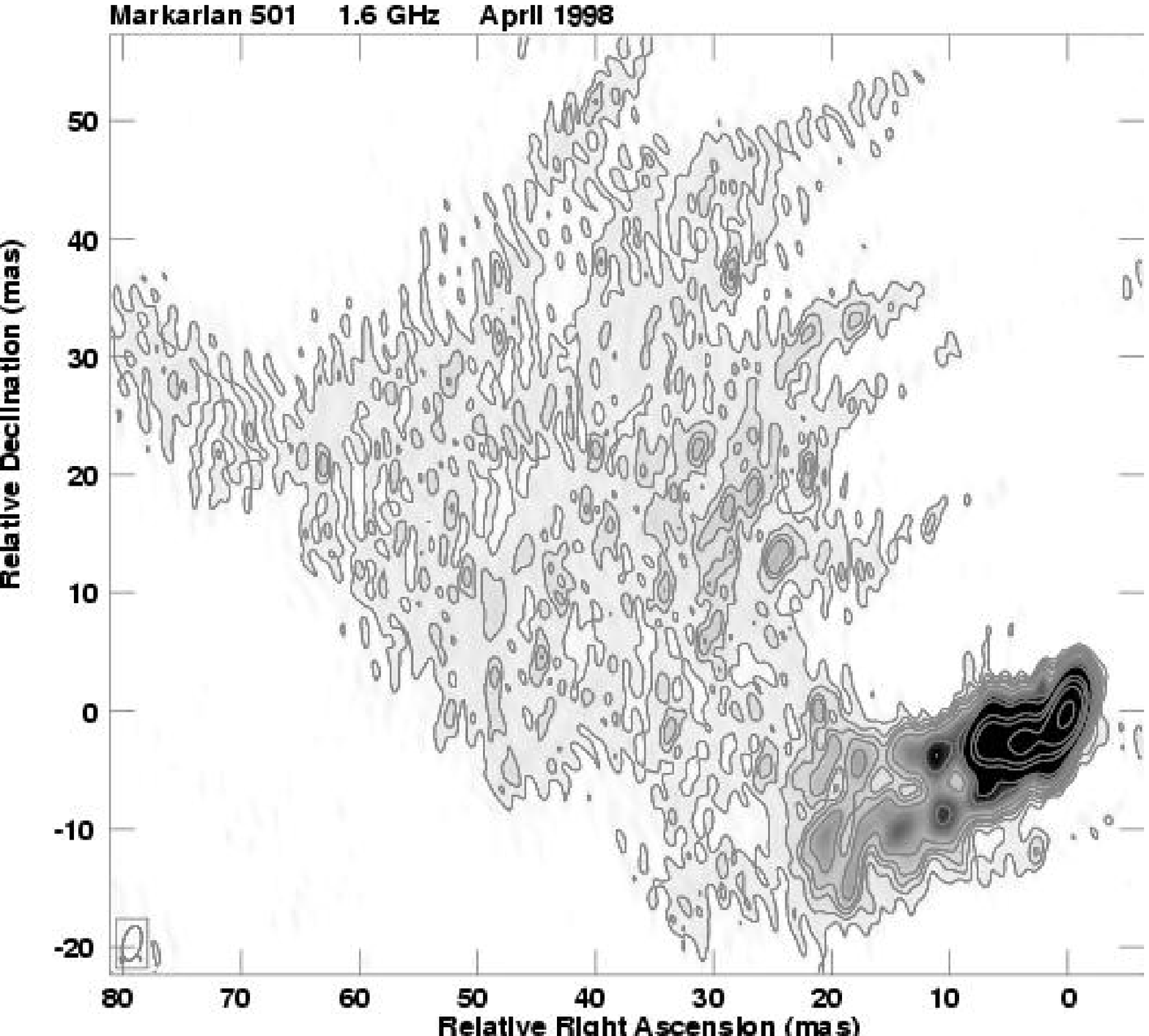}
\figcaption{Gray scale plus isocontour level of Mkn 501 from the 1.6\,GHz VSOP
observation in April 1998. The HPBW is 3\,$\times$\,1.5\,mas at PA $-10^\circ$. 
Contours are
drawn at 0.5, 1, 1.5, 2, 3, 5, 7, 10, 30, 50, 100, and 300 mJy/beam;
the greyscale range is 0--10\,mJy/beam. \label{fig8}}
\end{figure}

\begin{figure}
\plotone{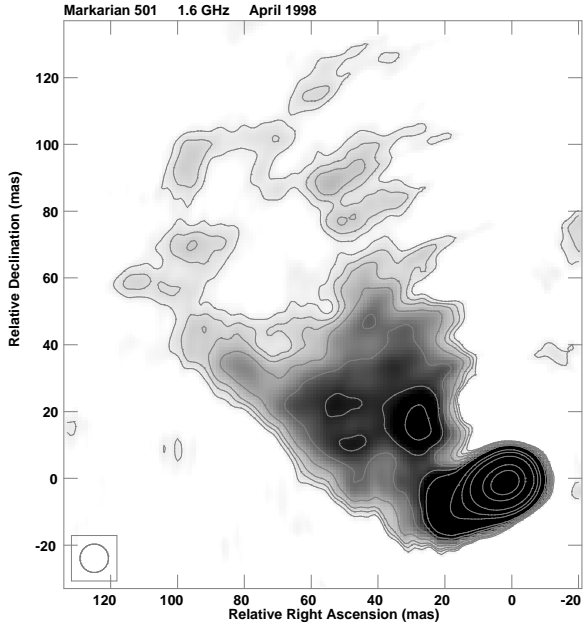}
\figcaption{Gray scale plus isocontour level of Mkn 501 from the 1.6\,GHz 
VSOP observation in April 1998. The image is convolved with circular beam 
of 8.5\,mas FWHM. The image noise level
is 0.4\,mJy/beam. Contours are drawn at 1, 1.5, 2, 3, 5, 7, 10, 30, 50,
100, 300, and 500\,mJy/beam; the greyscale range is 0.50--8\,mJy/beam.
\label{fig9}}
\end{figure}

\begin{figure}
\plotone{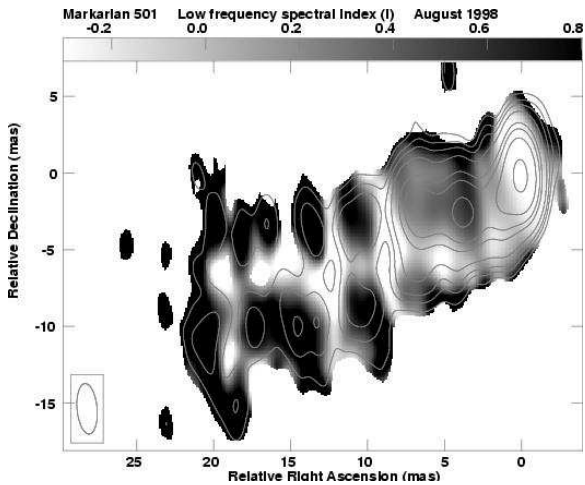}
\figcaption{Low frequency, low resolution spectral index image:
$\alpha_{1.6}^{4.8}$ with 3\,$\times$\,1.5\,mas HPBW (PA $10^\circ$). 
Contours show total intensity levels at 1.5, 3, 5, 10,
30, 50, 100, and 300\,mJy/beam (at\,1.6\,GHz); the greyscale is the 
spectral index from $-0.3$ (inverted) to 0.8 (steep). \label{spix1}}
\end{figure}

\begin{figure}
\plotone{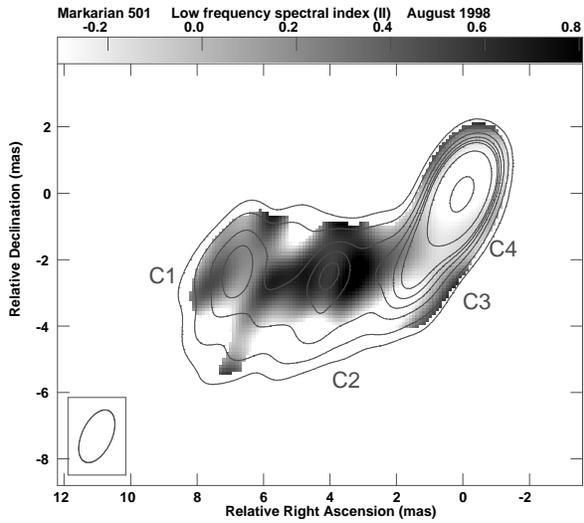}
\figcaption{Low frequency, high resolution spectral index image:
$\alpha_{1.6}^{4.8}$ with 2\,$\times$\,1\,mas HPBW (PA $-20^\circ$). 
Contours show total intensity levels at 5, 10, 20, 30,
40, 50, 100, 400 mJy/beam (at 4.8\,GHz); the greyscale is the spectral index
from $-0.3$ (inverted) to 0.8 (steep). \label{spix2}}
\end{figure}

\begin{figure}
\plotone{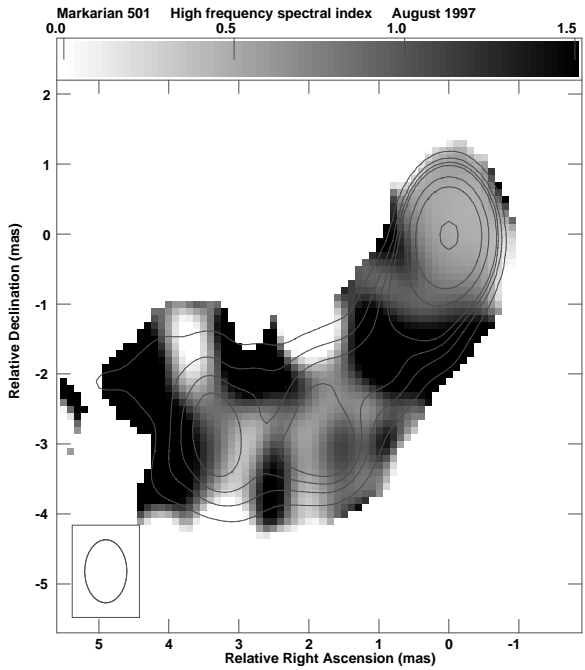}
\figcaption{High frequency, high resolution spectral index image:
$\alpha_{15}^{22}$ with $0.6 \times 0.9$\,mas HPBW
(RA\,$\times$\,Dec). Contours show total intensity levels at 5, 10, 15,
20, 50, 100, and 400\,mJy/beam (at 15\,GHz); the greyscale is the 
spectral index from 0 (flat) to 1.5 (steep), where $S(\nu) \propto
\nu^{-\alpha}$. \label{spix3}}
\end{figure}

The large scale radio structure has been imaged with the VLA
\citep{ulv83,van86,cas99}, showing a core dominated source with
two-sided diffuse emission oriented at $\mbox{PA} \sim 45^{\circ}$;
this indicates that the jets are non relativistic at a (projected)
distance of $\sim10$\,kpc from the core.  The pc-scale structure has
also been investigated, using VLBI techniques \citep[see
e.g.][]{con95}, with resolution of $\sim10$ mas. However, the
collection of multi-epoch, multi-frequency, high resolution and high
sensitivity images will improve our knowledge of the source morphology
and evolution.  The present VLBI observations show a strong core and a
one-sided jet. The jet exhibits multiple sharp bends before undergoing
a last turn, followed by rapid expansion. We can distinguish three
different regions:

\begin{itemize}

\item 
a first region, extending $\sim 10$\,mas from the core, where a high
brightness jet structure is present (Fig.~\ref{fig3}, \ref{fig4}, and
\ref{fig5}).  The jet PA is not constant in this region, being
150$^\circ$\,--\,160$^\circ$ near the core ($<$\,1\,mas) and moving to
$\sim 145^\circ$ (from 1 to $\sim$\,4\,mas from the core) to become
$\sim 90^\circ$ from 4 to 10\,mas.

The 22\,GHz high resolution image (Fig.~\ref{fig3}) shows that the jet
is resolved and limb-brightened starting at $\sim$\,1\,mas from the
core.  This is confirmed by the high sensitivity 5\,GHz Space VLBI
image (Fig.~\ref{fig5}); although this image does not have enough
angular resolution to resolve the inner jet near to the core, it shows
an extended low brightness emission, confirming the transverse
extension of the jet in the inner 2\,mas. In the region from 2 to
10\,mas from the core, the limb brightened jet structure is clearly
visible.

The brightness profile along the main axis of the jet is not uniform;
high brightness regions are followed by dimmer regions. In particular
at 5--6\,mas from the core a deep minimum is present just before an
extended bright spot with a peak 8--9\,mas from the core. However, we
do not refer to these regions as ``knots'', since they are resolved at
higher frequencies.

\item 
between $\sim$\,10\,mas and $\sim$\,30\,mas from the core, the jet
loses its collimation and becomes visible only in lower resolution
images (Fig.~\ref{fig6}, \ref{fig7}). In this region, the jet is
oriented at a PA of $\sim$ 110$^\circ$ and its opening angle is
increasing. A well defined limb brightened structure is visible (see,
e.g., Fig.~\ref{fig7}), in agreement with the ``layered'' structure
visible in the polarized emission \citep{aar99}. As in the first
region, lower frequency images reveal a wider jet and a more evident
external shear layer; this implies that the spectra in the inner
regions are flatter than in the external regions.

\item 
at $\sim$\,30\,mas from the core the jet shows strong bending and a
large opening angle (Fig.~\ref{fig8}, \ref{fig9}).  In this third
region the orientation of the diffuse jet is in agreement with the
orientation of the kpc-scale emission.  The jet opening angle is
$\sim$\,43$^\circ$. The radio brightness is uniform and there is no
indication of the helical structure suggested by \citet{con95}, which
possibly resulted from the poor $(u, v)$-coverage (five telescopes
only) of their observations.  A limb-brightened structure is visible
in our low resolution images up to $\sim$\,100\,mas from the core, as
can be seen in Fig.~\ref{fig9}.

\end{itemize}

No radio emission is detected either on the opposite side of the radio
core or in the position of the putative counterjet feature F reported
by \citet{con95}.  However, if we compare our measured total flux
densities to single dish data (e.g. from the UMRAO
database\footnote{see
\texttt{http://www.astro.lsa.umich.edu/obs/radiotel/umrao.html}}), the
correlated VLBI flux density accounts only for a part of the total
flux density ($\sim 70$\% at 5\,GHz and $\sim 65$\% at 15\,GHz); this
is in agreement with the presence of a larger scale structure resolved
out by the VLBI data, such as the two sided emission shown by
\citet{ulv83} and \citet{cas99}.

\subsection{Radio Flux Densities and Variability}
\label{sec:variability}

Single dish monitoring of Markarian 501 shows that this source is
remarkably stable in the radio band. \citet{ven01} monitored a sample
of 23 $\gamma-$ and X$-$ ray loud blazars over a 4 year period
(1996--2000) with the 32\,m Medicina radio telescope. From their data,
Markarian 501 is the most quiescent source, with an almost flat light
curve at both 5 and 8.4\,GHz. Moreover, Markarian 501 is part of the
UMRAO database, which also shows a fairly steady flux density. During
the period of our observation, the average flux densities are $(1.54
\pm 0.12)$\,Jy at 4.8\,GHz, $(1.53 \pm 0.13)$\,Jy at 8\,GHz and $(1.29
\pm 0.08)$\,Jy at 14.5\,GHz.

We consider the VLBI core flux density using our data at 15\,GHz
(0.6\,$\times$\,0.9 mas FWHM, RA\,$\times$\,Dec), where we have 8
different epochs (see Table~3). Assuming the map peak as the core flux
density, we see some dispersion in the values, which range between
360\,mJy (1997 Apr 25) and 508\,mJy (1995 Dec 15). This suggests
possible core variability, which is diluted in single dish
observations, since the VLBI core makes up only $\sim 30$\% of the
total flux density. We note, however, that our observations are not
homogeneous and were not aimed to monitor flux density variability.

\subsection{Proper Motion}
\label{sec:motion}

\begin{figure}
\plotone{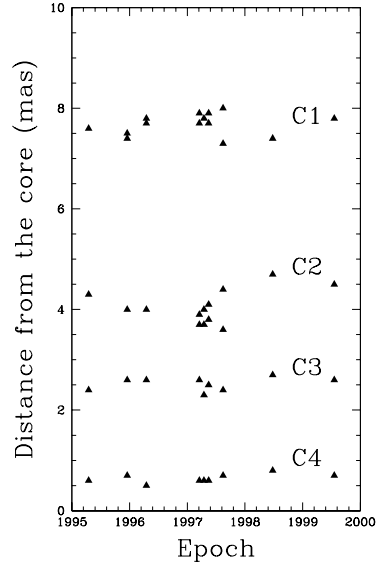}
\figcaption{Component positions from model-fitting. Distances are given
in milliarcsecond. All points are at 15\,GHz, with the exception of
the 1998.48 epoch which is at 8.4\,GHz. \label{positions}}
\end{figure}

The overall morphology of the source does not present dramatic changes
between epochs. No compact jet component (knot) is present in the jet
at all frequencies: if we consider the highest resolution data, the
local peaks present in the low resolution images become clearly
resolved. Therefore, we limit our study to the innermost region, where
high-resolution images are available (the jet brightness is too low to
be detected at high frequency beyond 10\,mas from the core). In this
region, the jet is always transversally resolved: it appears centrally
peaked at low resolution, but resolved with a limb brightened
structure at high resolution.  This implies that we could be mixing
together different components with different properties and moving at
different velocities. For all these reasons, we analysed images at the
best angular resolution (typically 0.6\,$\times$\,0.9\,mas,
RA\,$\times$\,Dec) and, in model-fitting, we tried to use a number of
components that did not produce an oversimplified representation of
the data. In particular, to avoid resolution or frequency effects, we
used only 15\,GHz data, plus the 8.4\,GHz observation at epoch
1998.48.  The results obtained from model-fitting of visibilities in
Difmap are presented in Table~4 and Fig.~\ref{positions}.  These
results are in agreement with those obtained by fitting Gaussian
components to the images (using \texttt{JMFIT} in AIPS).

The data are in general well fitted by a core and four jet components,
in agreement with \citeauthor{edw02} (not surprisingly, as most data
are the same). We labelled components according to the same notation
for ease of comparison: C1 to C4 moving from the outermost component
inward. However, for the reasons stated above, we adopted two
Gaussians for each of the components C2 and C1 at many epochs. This
produced a better fit to the data, in agreement with the resolved
limb-brightened jet structure present in these regions. Nominal
position errors are very small. The major uncertainty is related to
the complex source structure. We estimate a position error of
$\le$0.1\,mas for C4, C3, C2 and $\sim$\,0.2\,mas for C1.

From our analysis we conclude that no proper motion is visible in any
of the four components. Consider the case of component C2:
\citeauthor{edw02} measured a proper motion of $(0.6 \pm 0.1) c$ for
this component. However, we note that at the position of C2 two faint
external peaks are visible in the 22\,GHz image (Fig.~\ref{fig3}) on
either side of the lower frequency component. They could even have a
different spectrum (Fig.~\ref{spix1}). The peak position visible at
lower resolution in the 15\,GHz image is therefore the blend of the
limb-brightened structure visible at 22\,GHz. In the space VLBI image
at 5\,GHz, the component C2 is at a different position with respect to
the 15\,GHz images: this could be due to different spectra for the
different components.  We conclude that it is not possible to measure
a proper motion for such a complex structure and that any pattern
velocity measurement from low resolution data is not related to the
jet bulk speed in this source.

\subsection{Bulk Motion}
\label{sec:jet}

Under the assumption of intrinsically symmetric two-sided jets
affected by Doppler boosting effects, the observational data allow one
to estimate the jet bulk velocity.  In the following, we present the
results derived from a study of the jet/counterjet ratio, core
dominance and jet expansion.

\subsubsection{Jet/Counterjet Ratio}
\label{j/cj}

From the lack of detection of a counter-jet in all the images
(\S\,\ref{sec:morphology}), we set a lower limit on the jet/counterjet
ratio. We use the highest ratio between the jet brightness and the
counterjet upper limit. We estimate this ratio excluding the innermost
region of the jet; this avoids contamination from the unresolved
central emission. The best image for this analysis is the Space VLBI
observation obtained in Apr 1998 at 1.6\,GHz. From this image we can
derive a jet/counterjet ratio $R > 1250$ (1~rms) at $\sim$\,4\,mas
from the core. Given the well known relation $R =
\left(\frac{1+\beta\cos\theta}{1-\beta \cos\theta}\right)^{2+\alpha}$,
and assuming $\alpha = 0.5$ (see \S \ref{sec:index}), this value
yields $\beta \cos\theta > 0.89$. In turn, this implies $\beta
>$\,0.89 and $\theta <$\,27$^\circ$.

Our deep, low resolution images allow us to also put a constraint at
large distance from the core. In particular, in the 1.6\,GHz image
from the 2001.3 data, at $\sim 60$ mas from the core we still have R
$>$ 70; although less extreme, this value is also very interesting, as
it implies that the jet remains at least mildly relativistic ($\beta
\cos\theta > 0.69$) even at large distances from the core.

It is not possible to extend this analysis further with our
data. However, we remind the reader of the symmetric morphology of the
source on the kpc scale (\S\,\ref{sec:morphology}). This implies that
the bulk of the jet slows to subrelativistic speed going from the
parsec to the kpc scales.

\subsubsection{Core Dominance}

Given the general relation between total and core radio power
\citep{fer84,gio88,gio01,der90}, we can use low frequency unboosted
data to compute an ``expected'' value for the core power.  A
comparison between the expected and measured 5\,GHz core radio power
allows to constrain the jet orientation and velocity \citep{gio01}. We
note that at 5\,GHz self-absorption effects are small (see
\S\,\ref{sec:parameters}); however, to take also into account possible
core variability, we allow the observed core flux density to vary by a
factor of 2. This is a conservative assumption, based on the total
flux density monitoring and the range of derived VLBI core flux
densities (see \S\,\ref{sec:variability}), and, more generally, on
statistical flux density variability of AGN in the radio.

To estimate the orientation and jet velocity for Mkn 501, we have
compared its total unboosted flux density (1.81\,Jy) at 408\,MHz
\citep{fic85}, with the nuclear flux density at 5\,GHz. Since
\citet{gio01} used the arcsecond core radio power to derive their
correlation, we adopted the total VLBI correlated flux density at
5\,GHz ($\sim$\,800\,mJy) as the Mkn 501 core flux density. Any
possible underestimate in this assumption is compensated for by the
allowed range (a factor of 2) in flux density variability.  We found
that Mkn 501 has to be oriented at $\theta \le$\, 27$^\circ$ with
$\beta \ge$\,0.88. We note that a high jet velocity ($\beta >$\,0.95)
is allowed only in the range 10$^\circ$ $<$ $\theta$ $<$
27$^\circ$. High velocities are forbidden at smaller angles; the
resulting Doppler factor would require a core much brighter than
observed.

\subsubsection{Adiabatic Model}

The functional dependence of the jet intensity on the jet velocity and
radius for an adiabatically expanding jet has been discussed in the
case of relativistic motion by \citet{bau97}.  Since the model
represents a simplification of the real situation, the results should
be considered with caution and compared with other observational
results. This model has enabled useful constraints to be placed, in
agreement with other data, for 3C\,264 \citep{bau97} and other
sources, such as NGC\,315 \citep{cot99} and 3C\,449 \citep{fer99}.

To apply this model we used our deep VSOP observations at 1.6\,GHz
obtained in March 2001 and in April 1998. In these observations, the
jet emission is visible on scales up to $\sim 150$\,mas (see, e.g.,
Fig.~\ref{fig9}).  First, we derived brightness profiles across the
jet; we used the AIPS task \texttt{SLICE} on an image reconstructed
with a circular beam of 7\,mas diameter, where the longest baselines
have been omitted to increase the signal to noise ratio. In the
innermost region ($< 25$\,mas from the core), we have 8 slices, taken
every half beam-width at PA 27.2$^{\circ}$. After the main bend, we
considered 12 more slices, each one beam apart, at PA
$-57.1^{\circ}$. We stopped our analysis $\sim 100$\,mas from the
core; further out, the low jet brightness and the presence of extended
sub-structures do not allow the data to be fitted well.

\begin{figure}
\plotone{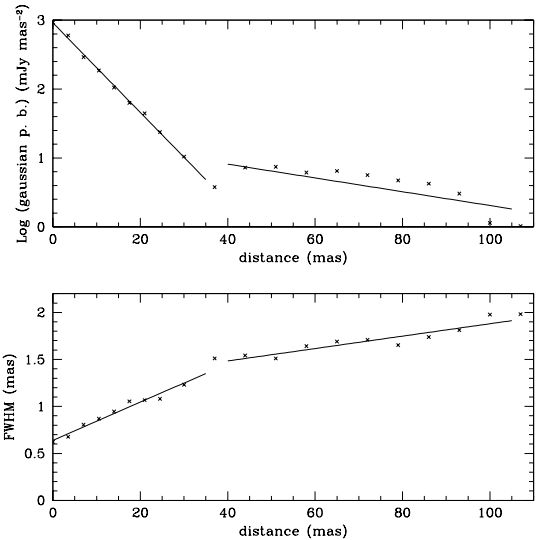}
\figcaption{Plots of peak brightness (top) and FWHM (bottom) of Gaussian
model fits vs.\ distance from the core. Straight lines represent
least-squares fits. There is a clear change in behavior at about 35\,mas
from the core. \label{trends1}}
\end{figure}

\begin{figure}
\plotone{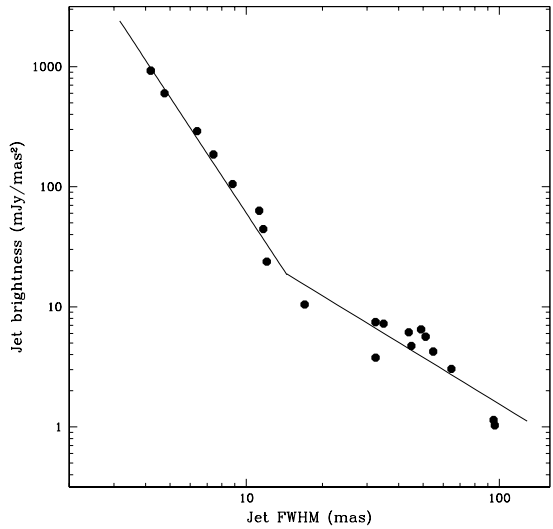} 
\figcaption{Plots of peak brightness vs.\ FWHM of Gaussian model
fits. Straight lines represent least-squares fits. The change in
behavior visible in Fig.~\ref{trends1} corresponds to the change in
slope at FWHM\,$\sim 15$\,mas.  The power law slopes are $-3.2$ and
$-1.3$. \label{trends2}}
\end{figure}

Using the AIPS task {\tt SLFIT}, we fitted single Gaussians to each
profile. This could be done unambiguously in most cases, although in a
few slices some deviation from a pure single Gaussian profile was
present. However, the difference between the area subtended by the
profile and the fit is always smaller than 5\% and does not affect the
analysis. We plot the resulting FWHMs and peak brightnesses in
Fig.~\ref{trends1} and \ref{trends2}; the quantities have been
deconvolved from the CLEAN beam, according to the formula given by
\citet{kil86}. There are clearly two regimes; least-squares fits yield
a power-law of index $-3.2$ in the inner jet ($<30$\,mas from the
core) and $-1.3$ in the outer regions. In the following, we use these
best-fit solutions rather then the actual data; this prevents us from
spurious results brought about by random fluctuations.

We recall the equations given in \citet{bau97} for the jet surface
brightness $I_{\nu}$:

$I_{\nu} \propto (\Gamma_j v_j )^{-(\delta +2)/3)} r_j^{-(5 \delta
+4)/3)} D^{2+\alpha} $ (predominantly parallel magnetic field)

$I_{\nu} \propto (\Gamma_j v_j )^{-(5 \delta +7)/6)} r_j^{-(7 \delta
+5)/6)} D^{2+\alpha} $ (predominantly transverse magnetic field).

\begin{figure}
\epsscale{0.45}
\plotone{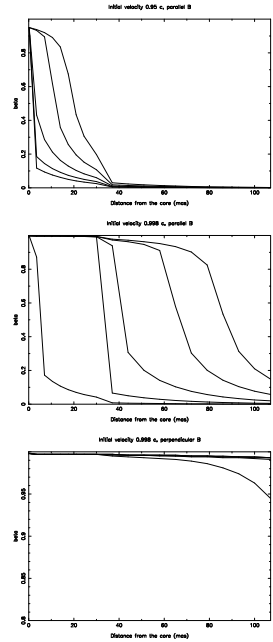}
\figcaption{Results of the adiabatic model fit. The low initial velocity with
parallel magnetic field is at top; middle and bottom plots show high
initial velocities ($\beta_i=0.998$) for a parallel and perpendicular
magnetic field, respectively. Orientation angles of $5^\circ,
10^\circ, 15^\circ, 20^\circ, 25^\circ$ are drawn in each plot, with
angle increasing from bottom left to top right. For illustration
purposes, the vertical axis in the bottom plot spans only the range
0.8--1. \label{adiabatic}}
\end{figure}

We assume an injection spectral index $\delta = 2$, as suggested from
the average $\alpha$ in the image, excluding the self-absorbed core
(\S\,\ref{sec:index}). It is not easy to decide which magnetic field
regime is preferable: according to the 8.4\,GHz images of
\citet{aar99}, the Mkn 501 jet is highly polarized at the edges with
the magnetic field parallel to the jet flow, while in the jet spine
the magnetic field is orthogonal to the jet flow with a lower degree
of order. \citet{pol03} report a 5\,GHz electric vector PA
distribution consistent with this result but the lower resolution
renders the results less conclusive.  Therefore, since the percentage
of ordered field and the intrinsic orientation are not well known, we
consider the two cases of pure parallel and perpendicular magnetic
field orientation; in a disordered magnetic field, the real result
will be in between the two extreme situations. Figure~\ref{adiabatic}
shows the derived trends of $\beta$ for an initial Lorentz factor
$\Gamma = 3.2$ and $\Gamma = 15$ ($\beta_i= 0.95, 0.998$) with
parallel and perpendicular magnetic fields. In each plot, we draw five
lines, corresponding to angles to the line of sight of 5$^{\circ}$,
10$^{\circ}$, ..., 25$^{\circ}$ (i.e., in the range of values allowed
by the jet sidedness and core dominance).

We note that the jet velocity decreases with the core distance, more
slowly in the perpendicular case, and on a shorter scale in the
parallel case.  The lack of detection of the CJ in all our images
strongly constrains any possible model since it implies a relativistic
jet also at large distance from the core ($> 50$\,pc). Models with low
initial Lorentz factor ($\Gamma < 5$) are definitely ruled out,
regardless of the assumed magnetic field orientation, since they
disagree with both the observed limb-brightened structure
(\S\,\ref{sec:jvs}) and the jet/counterjet ratio (at least in the case
of the parallel magnetic field); moreover, they require a jet
deceleration (\S\,\ref{sec:jvs}) between the $\gamma-$ray region and
radio jet region that is too strong.  Among models with a higher
initial speed ($\Gamma=15$), the case of the parallel magnetic field
requires an orientation angle of 25$^\circ$, at least at $>$\,70\,mas
from the core, to be in agreement with the observed jet/counterjet
brightness ratio. We cannot exclude models with a narrow angle to the
line of sight at small distances from the core, where solutions are
similar.  In the case of the perpendicular magnetic field, the
constraints are less severe since we have a fast jet even at large
distances from the core.

\subsection{Spectral Index}
\label{sec:index}

In Fig.~\ref{spix1}, \ref{spix2} and \ref{spix3}, we display the
spectral index distribution at various resolutions between 1.6 and
4.8\,GHz and between 15 and 22\,GHz. Observational results are
summarized in Table~5 for high resolution images and described in the
following subsections.

\subsubsection{Low Frequency Range: $\alpha^{1.6}_{4.8}$}

At low resolution ($3.3\,\times\,1.3$ mas, PA\,=\,$5^\circ$), we can
study the spectral index distribution at distances from the core
larger than 8--10 mas (Fig.~\ref{spix1}). The limb-brightened jet
structure between 8 and 20\,mas from the core is clearly visible. The
spectrum is always flat in the inner spine (medium value
$\alpha$\,=\,0.08) and steep in the bright shear regions, with values
in the range $0.9 \sim 2.1$. In the inner region ($<$\,8\,mas from the
core), the jet is well resolved with a steep spectrum in the more
external regions ($0.7 \sim 1.0$) but with a puzzling region south of
the high brightness region where the spectral index is $\alpha
\sim-0.1$. This region is clearly visible also in the lower resolution
image in \citet{edw00} where it seems connected to the inner spine
flat spectrum region, while no such connection is present in the total
intensity images.

Using the space VLBI baselines, we can study the source details at
higher angular resolution ($1.7 \times 0.9$\,mas, PA\,=\,$-25^\circ$)
even at this frequency (Fig.~\ref{spix2}). The nuclear spectrum is
inverted ($\alpha^{1.6}_{4.8} = -0.45$). Except for C4 (confused with
the core and subject to a large uncertainty), the jet sub-components
show flat to moderately steep spectra: $0.1 < \alpha^{1.6}_{4.8}
<0.6$.  Regions in between the components have steeper spectra.

\subsubsection{High Frequency Range: $\alpha^{15}_{22}$}

At these high frequencies, we are beyond the self-absorption turnover
and the nuclear region shows a flat, non-inverted spectrum
($\alpha$\,=\,0.45), generally steepening along the jet. The
resolution of 0.6\,$\times$\,0.9\,mas (RA\,$\times$\,Dec,
Fig.~\ref{spix3}) shows that the inner regions C4, C3, C2 have
$\alpha^{15}_{22}$ in the range $0.60 \sim 0.80$, while regions in
between are much steeper ($\alpha \sim2$ and higher). The large
component C1 is too faint and resolved to be properly imaged at these
high frequencies.  A marginal suggestion is present of a spectral
steepening moving from the inner (spine) to the external jet region
but at this high angular resolution the jet width is not well defined.

\subsubsection{Average Nuclear Spectrum}

It is interesting to consider the nuclear flux density over the whole
range of frequency. We are aware that the compact core of Mkn 501
presents some variability (see \S\ref{sec:variability}); however, the
large number of observations and frequencies available allow us to
carry out such a study based on average values.

\begin{figure}
\epsscale{1.00}
\plotone{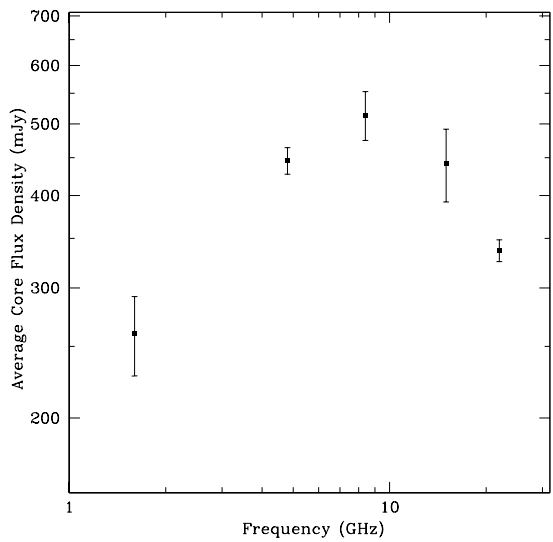}
\figcaption{Average core spectrum from data at resolution 
0.6\,$\times$\,0.9 (HPBW, RA\,$\times$\,Dec). Note the turnover at 
8.4\,GHz and the
rapid decrease at higher frequency. The low frequency part presents a
slow growth, suggesting the presence of more than one single
self-absorbed component. \label{spectrum}}
\end{figure}

We plot in Fig.~\ref{spectrum} the average core peak flux density,
from the images at resolution of 0.6\,$\times$\,0.9\,mas (see
Table~3); at 1.6\,GHz, we had to consider super-resolved images, in
order to avoid a large contamination from the jet. The spectrum is
convex, peaking at $\sim 8.4$\,GHz and steepening toward the high
frequency end. Below the turnover frequency, the spectrum is flatter
than $\alpha = 2.5$, as would be expected for a single self-absorbed
component. This suggests that different components with different
turnover frequencies may be blended together.

\section{DISCUSSION}
\label{sec:discussion}

\subsection{Jet Velocity}

\subsubsection{Velocity Structure}
\label{sec:jvs}

Limb brightened jet morphology on parsec scales is present in some
FR~I sources but also in a few high power FR~II sources
\citep{gio02}. A possible explanation is the presence of velocity
structure in the jet.  When viewed at a relatively large angle to the
line of sight, this structure yields different Doppler factors for the
internal spine and the external shear layer. In particular, an inner
high velocity spine could be de-boosted while a slower external layer
could be less strongly de-boosted, or even boosted.  Transverse jet
velocity structure has been found by \citet{lai99,lai02a,lai02b} in
3C\,31, where the jet velocity in outer regions is $\sim$\,0.7 the
inner spine velocity. \citet{chi00} invoke such a structure to account
for observational discrepancies between FR~I and BL-Lac objects.

A possible origin of the velocity structure is the interaction of the
jet with the surrounding medium and its consequent entrainment; the
mass loading slows down the jet in the external regions, while the
inner spine travels unaffected.  However, the presence of limb
brightened jets in both high and low power sources \citep{gio02} and
the evidence in M87 that the jet structure is present very near to the
core emission \citep[see the image presented by][]{jun99}, suggest the
possibility that the velocity structure may originate at the base of
the jet. \citet{mei03} presents a model where the inner high velocity
spine is produced in the black hole region, while the external shear
moving at lower velocity is produced in a more extended region
downstream.

In Mkn 501, it is interesting to note that the jet structure is
visible at a distance $<$\,1\,mas from the core, suggesting that it
originates very close to the base of the jet. However, we recall that,
at $z=0.034$, 1\,mas corresponds to $\sim0.7$\,pc, i.e., $>10^4$
Schwarzschild radii \citep[see][and references therein for a
discussion on the central black hole mass in Mkn 501]{rie03} and,
therefore, we do not have the angular resolution to see this jet inner
region. Nevertheless, we note that the presence of velocity structure
so near to the nuclear source implies too strong a deceleration if it
is completely due to the jet interaction with the ISM. We suggest
\citep[in agreement with M87 data of][]{jun99} that a transverse
velocity gradient is intrinsic to the jet structure \citep[as
discussed by][]{mei03} and that a further velocity decrease due to
mass loading in the jet from the ISM is likely to be present and
possibly dominates the jet velocity structure at larger distances from
the core.

\subsubsection{Bulk Motion and Jet Orientation}

We have strong indications that in Mkn 501 the radio jet emission is
oriented at a large angle with respect to the line of sight relative
to that expected for a blazar:

\begin{itemize}

\item
From the jet morphology: the jet is limb-brightened in high resolution
images at $\sim 1$\,mas from the core.  The different brightness can
be related to a different Doppler factor only with an orientation
angle $\theta$ \gtsim 15$^\circ$. Then, we can reproduce the observed
ratio between the brightness of the external shear and the inner spine
with $\Gamma_{spine} =15$, assuming $\Gamma_{layer} \sim 3$ (see
Tables~6, 7). With these values, the spine Doppler factor
($\delta_{spine}$) is lower than the external layer Doppler factor
($\delta_{layer}$).  On the contrary, a jet orientation at a smaller
angle and/or a lower spine velocity imply $\delta_{spine} >
\delta_{layer}$ and therefore a centrally peaked structure.

\item
From the core dominance: the core boosting is high but not extreme;
high bulk velocities require, therefore, an orientation angle $\theta$
in the range 10$^\circ$--27$^\circ$ with respect to the line of sight.

\item
From the fit to the trend of the jet brightness and FWHM: assuming an
adiabatic model and consistency with the jet/counterjet brightness
ratio, we need a radio jet starting with a high velocity ($\beta
\sim$\,0.998), and in the case of a parallel magnetic field, $\theta$
has to be $\sim$\,25$^\circ$ at $>$\,70\,mas.

\end{itemize}

These results seem to be in disagreement with constraints derived from
the high frequency ($\gamma$-ray) emission and variability detected in
blazars. Studying the rapid burst of TeV photons in Mkn 421,
\citet{sal98} determine that a jet with a bulk velocity
$\Gamma$\,\gtsim\,10 with opening angle and orientation angle $\sim
1/\Gamma$ is required. Similarly, in Mkn~501 the expected Doppler
factor should be \gtsim\,10, as obtained from fits of synchrotron
self-Compton (SSC) models \citep{kat01,tav01}.  To reconcile the radio
and high frequency results we have to assume that the inner region of
the jet (0.001--0.03\,pc) is moving with a Lorentz factor \gtsim\,10
and oriented at $\theta \, \ltsim \ 5^\circ$. From this region we have
high energy radiation and negligible radio emission.  Over the range
from 0.03 to 50\,pc (projected distance) the jet orientation has to
change from $\sim$\,5$^\circ$ to 25$^\circ$ (or at least $\sim
15^\circ$ in the case of perpendicular magnetic field).

We propose that this large change in the jet direction may be gradual
and we derive the minimum change requested by observational data at
different distances from the core. As shown in Table~6, a progressive
change in the jet direction with respect to the line of sight is
possible and in agreement with the observed pc-scale jet properties.

At present, there is no obvious explanation for this change in the jet
direction in the inner region.  It is possible that it could be
related to helical type motion \citep{vil99} (although helical motion
is not visible in our high resolution images), or to an instability in
the jet due to reasons that are thus far unknown.  It is also
important to stress that small changes in direction are not uncommon
in FR~I radio galaxies and are magnified when projection effects
occur. In any case, we recall that beyond 30\,mas from the core the
jet position angle becomes stable and closely aligned with the
kpc-scale structure.

We note that in the case of the perpendicular magnetic field the jet
velocity is still very high at large distances from the core, in
agreement with the lack of detection of a visible counterjet in all
published maps of this source and despite the symmetric structure
visible in the VLA low resolution image \citep{ulv83}. However, we
recall that $\theta \ \gtsim \ 15^\circ$ is required to justify the
observed limb-brightened structure from Doppler boosting alone, even
in the case of perpendicular magnetic field.

\subsection{Physical Parameters}
\label{sec:parameters}

From the present data we can derive that:

\begin{itemize}

\item
The nuclear region shows a self-absorbed spectrum with a peak
frequency $\nu_m \sim 8.4$\,GHz and a peak flux density $S_m \sim
0.55$\,Jy.  Following \citet{mar87}, we can use these data to estimate
the magnetic field in the nuclear region (from 0.03 to 0.15\,pc):

B = $3.2 \times 10^{-5} \, \theta^4 \, \nu^5_m \, S^{-2}_m \, \delta
\, (1+z)^{-1};$

assuming $\theta$ (the core angular size) = 0.2\,mas (0.15\,pc) and
$\delta$\,=\,5 (Table~6), we estimate a magnetic field
B\,$\sim$\,0.03~gauss.  On the other hand, fits of SSC models to a
high energy state observed in April 1997 yield values of the magnetic
field in the range 0.2--0.02~gauss in the inner region ($<$\,0.03\,pc)
\citep{kat01}. This suggests a constant or a not very large decrease
of the magnetic field from the gamma ray emitting region to the
beginning of the radio jet.

\item
The radio spectra of jet sub-components show an evident steepening at
high frequencies, suggesting the effect of radiative losses in a high
magnetic field.

\item
In the region where the jet is well resolved, the spectrum is flat in
the inner spine confirming the presence of a higher transport
efficiency, lower radiative losses, or reacceleration mechanisms. The
transverse steepening from the spine to the shear layer regions
confirms the larger losses in these regions, probably related to the
interaction with the external medium.

\item
It is not straightforward to determine the equipartition magnetic
field \citep{pac70} given the complex structure of the jet. We can
obtain a rough estimate by assuming that electrons and heavy particles
carry the same amount of energy ($K = 1$), the filling factor
$\phi$\,=\,1, and the frequency range is between 10\,MHz and
100\,GHz. This yields $\sim$\,0.015~gauss in the C4 and C3 regions and
0.01~gauss in the C2 and C1 regions.

\item
Assuming a viewing angle $\theta$ as in Tables~6 and 7, the observed
angular distances correspond to the following de-projected linear
distances:

-- The inner limb-brightened region is present at $<$\,4\,pc from the
   core

-- The first jet region extends to $\sim$\,10\,mas, corresponding to 
   $\sim$\,30\,pc

-- The VLBI jet is visible up to 170--270\,pc (assuming $\theta$ =
   25$^\circ$ -- 15$^\circ$, respectively)

-- The kpc-scale structure is symmetric (i.e. not relativistic) at
   $\le 20 - 30$\,kpc from the core

-- The large scale symmetric structure extending to $\sim$\,70$''$
   corresponds to a linear size of 120--200\,kpc, in agreement with
   the expected linear size of a FR I radio galaxy with the same low
   frequency total radio power.

\end{itemize}

\section{CONCLUSIONS}
\label{sec:conclusions}

As a result of the many frequencies and resolutions available, we have
established that a limb-brightened structure is indeed present in
Markarian 501, beginning in the very inner jet. This has important
consequences: (1) it implies the presence of velocity structure
already at very small scale, which jet models will have to take into
account; (2) it suggests, together with other arguments, that the VLBI
jet can not be oriented at too small a viewing angle. In particular,
the detection of a very long one-sided jet and the results of an
adiabatic expansion fit yield values of the angle of view in the range
$15^\circ < \theta < 25^\circ$, depending also on the magnetic field
orientation.

The bulk velocity is strongly relativistic at the beginning ($\Gamma
\sim 15$), decreasing to $\beta \sim 0.6$ ($\Gamma \sim$\,1.25) at the
end of the VLBI jet, 100\,mas from the core (parallel $B$) or to
$\beta \sim 0.99$ ($\Gamma \sim 7$) in the case of a perpendicular
$B$. The data considered here are consistent with a pattern speed of
zero. Previous reports, based mostly on lower frequency observations,
have reported low pattern speeds, however the complex limb-brightened
structure revealed at 22\,GHz makes measurement of component speeds
problematic at high frequencies.

Finally, the study of the spectral indices has revealed a
self-absorbed compact core and a complex spectral index distribution
in the jet with flat and very steep regions. A clear connection is
present between limb-brightening in total intensity and the spectral
index images: the inner spine has a flat spectrum while the jet more
external regions show steep spectra.  Our derived magnetic field
strengths are similar to those obtained from data at other
wavelengths, yielding values of order $B \sim 0.03$~gauss in the core
region. The estimated equipartition field in the jet region is in the
range 0.015--0.01~gauss.

\begin{acknowledgments} 
The authors thanks R.\ Fanti for his critical reading and useful
comments on the paper. MG thanks JIVE (Joint Institute for VLBI in
Europe) for support and hospitality and D.\ Gabuzda for helpful advice
in the reduction of the third epoch of VSOP data. We gratefully
acknowledge the VSOP Project, which is led by the Japanese Institute
of Space and Astronautical Science in cooperation with many
organizations and radio telescopes around the world. The National
Radio Astronomy Observatory is operated by Associated Universities,
Inc., under cooperative agreement with the National Science
Foundation. This research has made use of NASA's Astrophysics Data
System Bibliographic Services and of the United States Naval
Observatory (USNO) Radio Reference Frame Image Database (RRFID). This
research has made use of data from the University of Michigan Radio
Astronomy Observatory which is supported by funds from the University
of Michigan.  This material is based in part upon work supported by
the Italian Ministry for University and Research (MIUR) under grant
COFIN 2001-02-8773 and by the U.S.\ National Science Foundation under
Grant No.\ 0098579.

\end{acknowledgments}

\clearpage

\begin{deluxetable}{lclc}

\tablecaption{Observation List \label{table1}}
\tablehead{
\colhead{Date} & \colhead{Frequency} & \colhead{Array} &
\colhead{Observing Time} \\
\colhead{} & \colhead{(GHz)} & \colhead{} &  \colhead{(min)}}

\startdata

1995 Apr 7\tablenotemark{a}	& 15		& VLBA		& 40 \\
1995 Apr 12\tablenotemark{b} 	& 8.3		& VLBA		& 15 \\
1995 Dec 15\tablenotemark{a}	& 15		& VLBA		& 40 \\
1996 Apr 23\tablenotemark{b} 	& 8.3		& VLBA		& 7.5 \\
1996 Apr 23	& 15		& VLBA		& 6 \\
1996 Jun 7\tablenotemark{c}	& 8.1		& VLBA		& 6 \\
1997 Mar 13\tablenotemark{a}	& 15		& VLBA		& 40 \\
1997 Apr 25\tablenotemark{d}	& 15		& VLBA		& 90 \\
1997 Apr 25     & 22            & VLBA          & 90 \\
1997 May 26\tablenotemark{d}	& 15		& VLBA		& 50 \\
1997 May 26     & 22            & VLBA          & 50 \\
1997 Aug 4	& 1.6		& VLBA$-$SC+GO+{\bf HALCA}& 420	\\
1997 Aug 15\tablenotemark{d}	& 15		& VLBA		& 40 \\
1997 Aug 15     & 22            & VLBA          & 40 \\
1998 Apr 7	& 4.8		& VLBA+EB+{\bf HALCA}	& 780 \\
1998 Apr 8	& 1.6		& VLBA+RO+GO+{\bf HALCA}& 600 \\
1998 Jun 24\tablenotemark{b}	& 8.4		& VLBA+GC+GN+KK+ & 20 \\
		&		& MC+ON+WF	&\\
1999 Jul 19\tablenotemark{a}	& 15		& VLBA$-$SC	& 40 \\
2001 Mar 5	& 1.6		& VLBA+GO+RO+{\bf HALCA}& 480 \\
2001 Mar 6	& 4.8		& VLBA$-$HN+{\bf HALCA}	& 540 \\

\enddata

\tablecomments{EB, Effelsberg (Germany) 100~m, GC, Gilcreek (USA) 26~m
GN, Green Bank (USA) 20~m; GO, Goldstone (USA) DSN 70~m; HN, Hancock VLBA
25~m; KK, Kokee Park (USA) 20~m; MC, Medicina (Italy) 32~m; ON, Onsala
(Sweden) 20~m; RO, Robledo (Spain) DSN 70~m; SC, St.\ Croix VLBA 25~m;
WF, Westford (USA) 18~m.}

\tablenotetext{a}{VLBA 2\,cm Survey of Compact Radio Sources \citep{kel98}}
\tablenotetext{b}{RRFID: Radio Reference Frame Image Database
\citep{fey97}}
\tablenotetext{c}{ICRF: International Celestial Reference Frame
\citep{ma98}} 
\tablenotetext{d}{VLBA Project BM082 \citep{mar99}}

\end{deluxetable}

\clearpage

\begin{deluxetable}{lccll}
\tablecaption{Summary of VLBA+HALCA Observations \label{tab:vsop_obs}}
\tablehead{
\colhead{Date}	& \colhead{Frequency}	& \colhead{Obs. Time}	& \colhead{Tracking} 	& \colhead{Other} \\
\colhead{}	& \colhead{(GHz)}	& \colhead{(hr)}	&
\colhead{Stations} 	& \colhead{Telescopes}}

\startdata

1997 Aug 4	& 1.6	& 7		& NZ ($3^h$) 	& GO ($4^h$) \\ 
1998 Apr 7	& 4.8	& 13		& RZ ($2.7^h$+$1.1^h$), NZ ($3.3^h$) & EB \\
1998 Apr 8	& 1.6	& 10		& RZ ($3.6^h$), NZ ($4^h$) &
GO, RO \\
2001 Mar 5 	& 1.6	& 8		& RZ ($1^h$), NZ ($2^h$) & RO
($4^h$), GO ($7.5^h$) \\
2001 Mar 6	& 4.8	& 9		& RZ ($0.5^h$), NZ ($1^h$)
&  \nodata \\

\enddata

\tablecomments{RZ, Robledo tracking station; NZ, Green Bank tracking
station; EB, Effelsberg (100\,m); RO, Robledo (70\,m); GO, Goldstone
(70\,m).}

\end{deluxetable}

\clearpage

\begin{deluxetable}{lclcc}
\tablecaption{Observational Parameters \label{tab:parameters}}
\tablehead{
\colhead{Date} & \colhead{Frequency} & \colhead{HPBW\tablenotemark{a}}
& \colhead{Noise Level} & \colhead{Peak Flux Density} \\
\colhead{}     & \colhead{(GHz)}     & \colhead{(mas)} & \colhead{(mJy/beam)}  & \colhead{(mJy/beam)}}

\startdata

1995 Apr 7	& 15		& 0.6 $\times$ 0.9	& 0.20        & 485 \\
		&   		& 1.2			& 0.25        & 524 \\
1995 Apr 12 	& 8.3		& 1.2			& 0.40        & 524 \\
1995 Dec 15	& 15		& 0.6 $\times$ 0.9	& 0.35        & 508 \\
		&   		& 1.2			& 0.30	      & 546 \\
1996 Apr 23	& 8.3		& 1.2			& 0.25        & 521 \\
		& 15		& 0.6 $\times$ 0.9	& 0.30 	      & 490 \\
 		&   		& 1.2   		& 0.35        & 520 \\
1996 Jun 7	& 8.1		& 1.2			& 0.40        & 535 \\
1997 Mar 13	& 15		& 0.6 $\times$ 0.9	& 0.15        & 381 \\
		&   		& 1.2			& 0.20        & 410 \\
1997 Apr 25	& 15		& 0.6 $\times$ 0.9	& 0.10        & 360 \\
 		&   		& 1.2			& 0.15 	      & 390 \\
	        & 22            & 0.3 $\times$ 0.5      & 0.15        & 327 \\
1997 May 26	& 15		& 0.6 $\times$ 0.9	& 0.10	      & 408 \\
 		&   		& 1.2			& 0.10	      & 438 \\
	        & 22            & 0.4 $\times$ 0.6      & 0.20 	      & 334 \\
1997 Aug 4	& 1.6		& 1.5 $\times$ 2.9	& 2.3         & 409 \\
1997 Aug 15	& 15		& 0.6 $\times$ 0.9	& 0.20        & 455 \\
		&   		& 1.2			& 0.15        & 487 \\
	        & 22            & 0.3 $\times$ 0.5      & 0.25        & 350 \\
1998 Apr 7	& 4.8		& 0.6 $\times$ 0.9 	& 0.40        & 465 \\
		&    		& 1.2    		& 0.25        & 509 \\
1998 Apr 8	& 1.6		& 1.5 $\times$ 2.9      & 0.12        & 421 \\
1998 Jun 24	& 8.6		& 0.6 $\times$ 0.9    	& 0.25        & 477 \\
		&    		& 1.2           	& 0.35	      & 508 \\
1999 Jul 19	& 15		& 0.6 $\times$ 0.9	& 0.20	      & 446 \\
		&   		& 1.2    		& 0.20	      & 480 \\
2001 Mar 5	& 1.6		& 1.5 $\times$ 2.9      & 0.50	      & 316 \\
2001 Mar 6	& 4.8		& 1.2           	& 0.50        & 428 
\enddata

\tablenotetext{a}{Column 3: circular beam or RA\,$\times$\,Dec in PA 0$^{\circ}$}

\end{deluxetable}

\clearpage

\begin{deluxetable}{lcr}
\tabletypesize{\footnotesize}
\tablecaption{Results of the Model - Fitting \label{tab:modelfit}} 
\tablehead{
\colhead{Epoch} & \colhead{Component} & \colhead{Distance from core}
\\
\colhead{}      & \colhead{}          & \colhead{(mas)}}

\startdata
1995.29& C4   & 0.6          \\
       & C3   & 2.4          \\
       & C2   & 4.3          \\
       & C1   & 7.6          \\
1995.96& C4   & 0.7          \\
       & C3   & 2.6          \\
       & C2   & 4.0 \\
       & C1a  & 7.4 \\
       & C1b  & 7.5 \\
1996.29& C4   & 0.5 \\
       & C3   & 2.6 \\
       & C2   & 4.0 \\
       & C1a  & 7.7 \\
       & C1b  & 7.8 \\
1997.21& C4   & 0.6 \\
       & C3   & 2.6 \\
       & C2a  & 3.7 \\
       & C2b  & 3.9 \\
       & C1a  & 7.7 \\
       & C1b  & 7.9 \\
1997.29& C4   & 0.6 \\
       & C3   & 2.3 \\
       & C2a  & 3.7 \\
       & C2b  & 4.0 \\
       & C1   & 7.8 \\
1997.37& C4   & 0.6 \\
       & C3   & 2.5 \\
       & C2a  & 3.8 \\
       & C2b  & 4.1 \\
       & C1a  & 7.7 \\
       & C1b  & 7.9 \\
1997.62& C4   & 0.7 \\
       & C3   & 2.4 \\
       & C2a  & 3.6 \\
       & C2b  & 4.4 \\
       & C1a  & 7.3 \\
       & C1b  & 8.0 \\
1998.48& C4   & 0.8 \\
       & C3   & 2.7 \\
       & C2   & 4.7 \\
       & C1   & 7.4 \\
1999.55& C4   & 0.7 \\
       & C3   & 2.6 \\
       & C2   & 4.5 \\
       & C1a  & 7.8 \\
       & C1b  & 8.9
\enddata

\end{deluxetable}

\clearpage

\begin{deluxetable}{lccrrrrrrrr}
\tablecaption{Spectral Index Results \label{tab:spectrum}}
\tablehead{
\colhead{Image} & \colhead{beam} & \colhead{PA} & \colhead{core} & \colhead{C4} & \colhead{C4toC3} & \colhead{C3} & \colhead{C3toC2} & \colhead{C2} & \colhead{C2toC1} & \colhead{C1}}

\startdata

$\alpha^{1.6}_{4.8}$ & $1.7 \times 0.9$ & $-25^\circ$ & $-0.45$ & $-0.2^*$ & $0.6^*$ & 0.1  &  0.9   & 0.6  &   0.7  & 0.3 \\
$\alpha^{15}_{22}$   & $0.6 \times 0.9$ & $0^\circ$   & 0.45    & 0.80     & 2       & 0.65 & \nodata & 0.70 & 2.4    & resolved \\

\enddata

\tablecomments{Results of the spectral index imaged in Figures~\ref{spix2}, 
\ref{spix3}. Typical uncertainties are $\pm 0.05$. Data
marked with an asterisk (*) are estimated after core subtraction and have
larger uncertainties.}

\end{deluxetable}

\clearpage

\begin{deluxetable}{clllllc}
\tablenum{6}
\tablecaption{Jet Velocity Structure - Parallel Magnetic Field \label{tab:6}}
\tablehead{
\colhead{R$_{core}$} & \colhead{$\theta$} & \colhead{$\Gamma_{spine}$} & \colhead{$\delta_{spine}$} & \colhead{$\Gamma_{layer}$} & \colhead{$\delta_{layer}$} & \colhead{Notes}  \\
\colhead{(pc)}       & \colhead{($^\circ$)} & \colhead{} & \colhead{} & \colhead{} & \colhead{} & \colhead{} }

\startdata
0.0001 -- $<$0.03 & 4    &   15             &   15             &  ?               &  ?               &$\gamma$-ray region \\
0.03 -- 0.15   & 10   &   15             &   4              &  10              & 5                & Radio core \\
0.15 -- 7      & 15   &   15             &   2              &  3               & 5                & First jet region \\
7 -- 20        & 15--20&  15             & 2--1             & 3                & 4--3             & Before of large bending\\
20--30         & 25   & 10--3            & 1--2             & 2                & 2.5              & After the large bending\\
50             & 25   & 1.25             & 1.8              & 1.1                & 1.5                & Final VLBI jet region 
\enddata

\tablecomments{R$_{core}$ = projected distance from the core}

\end{deluxetable}

\begin{deluxetable}{clllllc}
\tablenum{7}
\tablecaption{Jet Velocity Structure - Perpendicular Magnetic Field \label{tab:7}}
\tablehead{
\colhead{R$_{core}$} & \colhead{$\theta$} & \colhead{$\Gamma_{spine}$} & \colhead{$\delta_{spine}$} & \colhead{$\Gamma_{layer}$} & \colhead{$\delta_{layer}$} & \colhead{Notes}  \\
\colhead{(pc)}       & \colhead{($^\circ$)}}

\startdata
0.0001 -- $<$0.03 & 4    &   15             &   15             &  ?               &  ?               &$\gamma$-ray region \\
0.03 -- 0.15   & 10   &   15             &   4              &  10              & 5                & Radio core \\
0.15 -- 7      & 15   &   15             &   2              &  3               & 5                & First jet region \\
7 -- 20        & 15   &  15             & 2                 & 3                & 4             & Before of large bending\\
20--30         & 15   & 10              & 2.5             & 3                & 4              & After the large bending\\
50             & 15   & 10             & 2.5              & 3              & 4                & Final VLBI jet region 
\enddata

\tablecomments{R$_{core}$ = projected distance from the core}

\end{deluxetable}

\clearpage

\end{document}